# Macroeconomic evaluation of the growth of the UK economy over the period 2000 to 2019

Laurence Francis Lacey, Skerries, County Dublin, Ireland

03 Nov 2022

## Abstract

An information entropy statistical methodology was used to evaluate the growth of the UK economy over the period 2000 to 2019, with an emphasis on the impact of labour productivity on gross domestic product (GDP) per capita and the average growth in real wages, during this time period.

The growth of the UK economy over the period 2000 to 2019 can be described in terms of three distinct phases:

1) 2000 to 2007 - strong sustained economic growth
2) 2008 to 2013 - the impact of the international financial crisis, its immediate aftermath, and period of recovery
3) 2014 to 2019 - weak sustained economic growth

The key determinant of the UK economic performance over this period would appear to the annual rate of growth in labour productivity. It was closely related to the annual rate of growth in GDP per capita, and it was significantly weaker in the period 2014 to 2019 compared to the period 2000 to 2007. This also corresponded with a weaker rate of growth in annual average real wages over the period 2014 to 2019 compared to the period 2000 to 2007. Throughout the period 2000 to 2019, UK CPI was maintained, on average, at approximately 2.1% per annum.

More rapid UK economic growth would be expected to be achieved by sustained investment in measures that enhance labour productivity, with the further expectation that a sustained improvement in labour productivity would increase the annual rate of growth of UK GDP per capita and average real wages. While the results given in this paper are specific to the UK over the time period 2000 to 2019, the expectation is that the methodology and approach adopted can be applied to quantifying the dynamics of any developed economy over any time period.

## 1. Introduction

The objectives of this paper were to investigate the nature of the growth of the United Kingdom (UK) economy over the time period 2000 to 2019 and, in particular, to explore the quantitative relationships between the growth in:

1) UK GDP per capita and UK labour productivity, and
2) Average UK real wages and UK labour productivity

The time period 2000 to 2019 was chosen, because in the middle of this time period, the international financial crisis of 2007 to 2009 occurred [1] (also known as the "Great Recession" [2]).

# 2. Methods

## 2.1 Statistical Methodology

An information entropy statistical methodology has been developed for investigating expansionary processes, in which full details of the methodology can be found [3,4,5]. This was the methodology that was used for the macroeconomic evaluation undertaken in this paper.

For an exponential expansionary process, the methodology provides a rate-constant (λ) for the exponential growth of the process (G(t)) and the associated information entropy (IE) for the time series under investigation [3,5]. This can be expressed as follows:

$$\mathrm{G(t)} = \exp(\lambda\, x\, \mathrm{t})$$

and,

$$\mathrm{IE\ G(t)} = \lambda\, x\, \mathrm{t}$$

with,

$$\mathrm{IE\ G(t = 0)} = 0$$

While information entropy has no units, at any given time, it is related to the growth rate ($r$), where:

$$r = \exp(\lambda) - 1 \quad \text{and} \quad \lambda = log_e(1 + r)$$

## 2.2 Time Series Data

The sources of the time series data, over the period 2000 to 2019, are: UK Gross Domestic Product (GDP) [6], UK Consumer Price Index (CPI) [7], UK GDP per capita [8], UK labour productivity [9], and average UK real wages [10].

All plots of the data analysis given below were obtained using Microsoft Excel 2019, 32-bit version.

# 3. Results

## 3.1 Growth of UK GDP and CPI over the period 2000 to 2019

The growths of UK GDP and CPI, relative to the year 2000, over the period 2000 to 2019 are shown in Figure 1. The information entropy transformed (IE) UK GDP and IE UK CPI, relative to the year 2000, over the period 2000 to 2019 are shown in Figure 2. The rate of growth in UK GDP following the international financial crisis of 2007 to 2009 is lower than the years preceding that event (Figure 2a). Throughout the period 2000 to 2019, UK CPI was maintained, on average, at approximately 2.1% per annum (Figure 2b).

Figure 1: Growth of UK GDP and CPI, relative to the year 2000, over the period 2000 to 2019

(a) Growth of UK GDP over the period 2000 to 2019

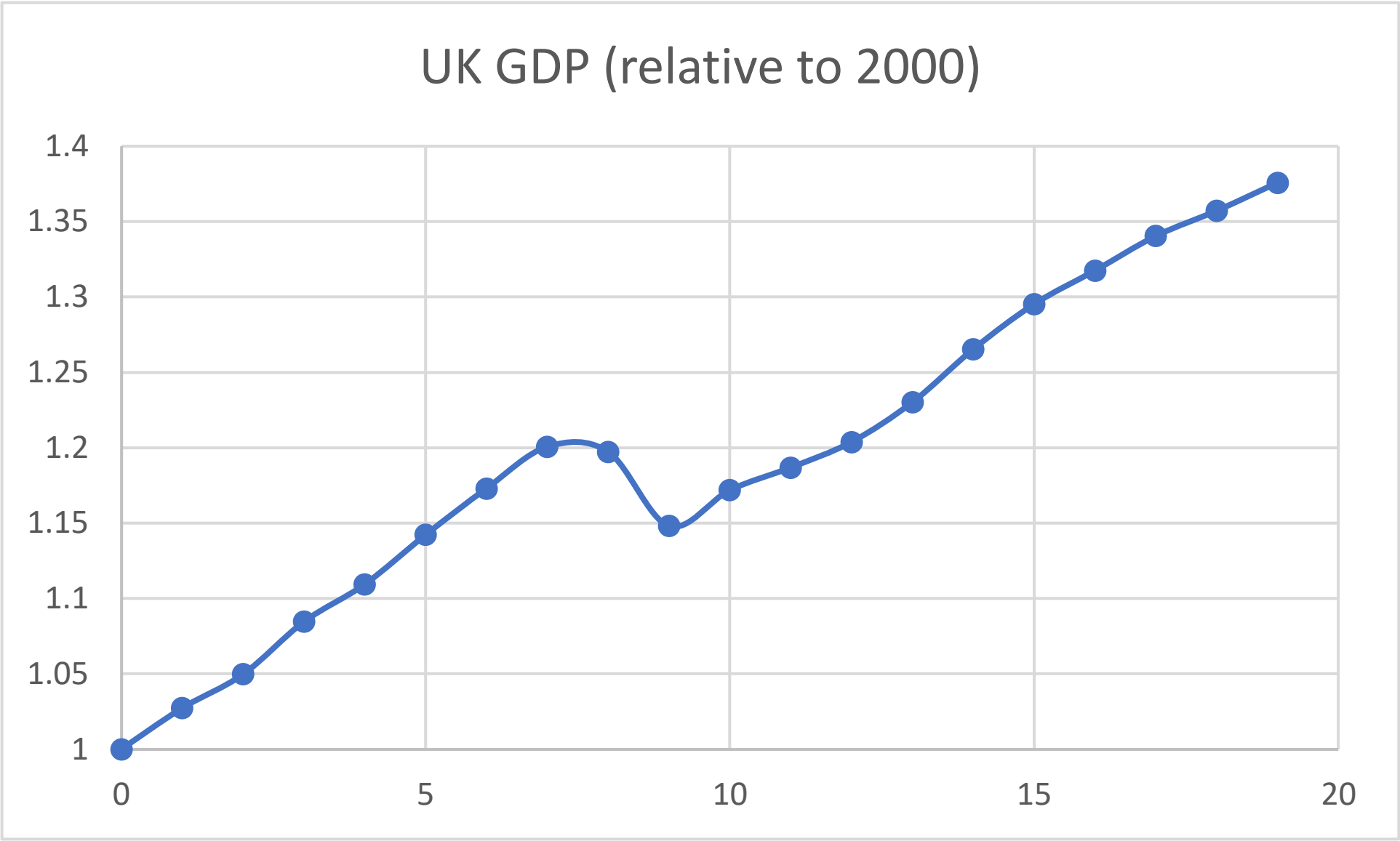

(b) Growth of UK CPI over the period 2000 to 2019

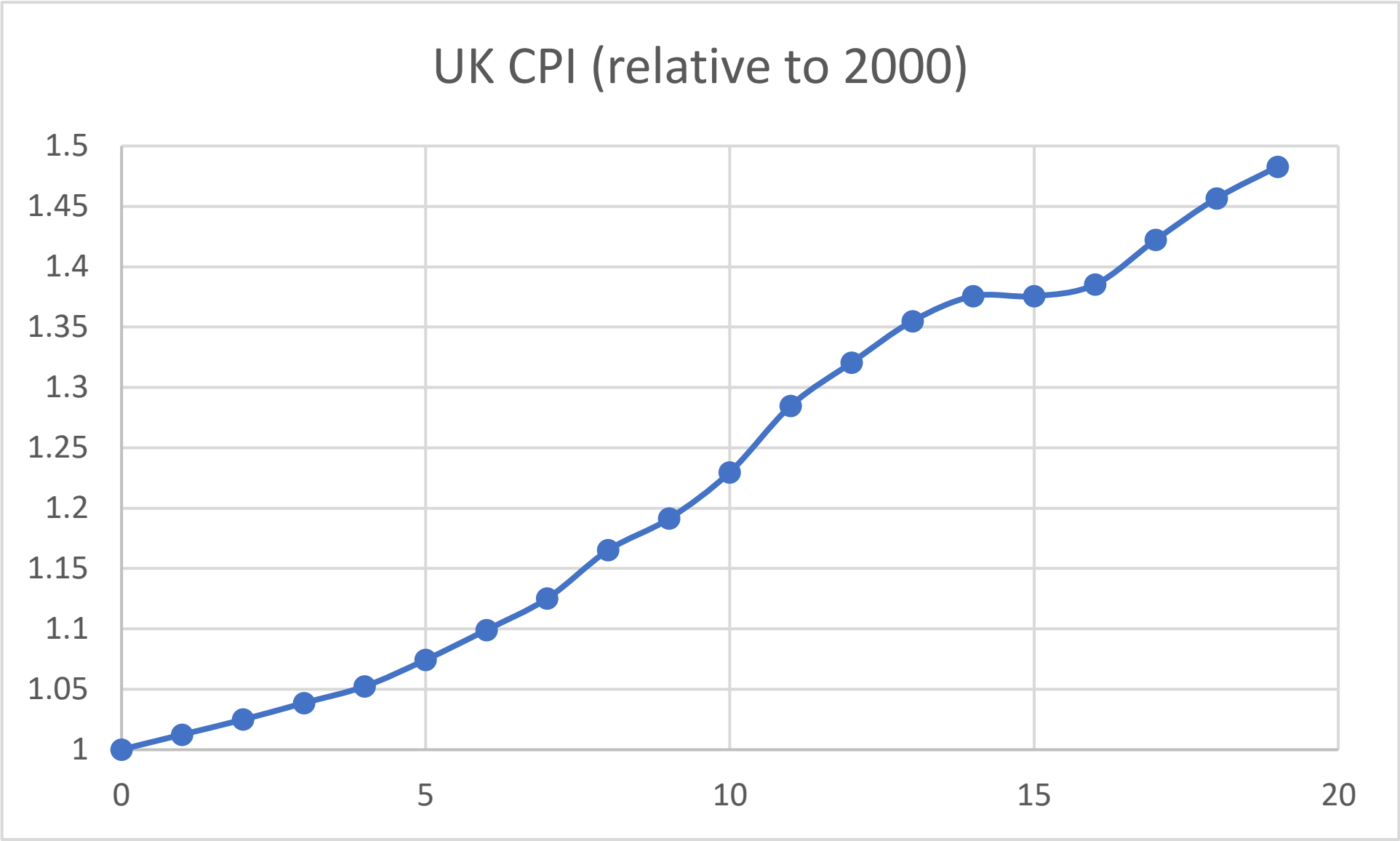

Note: UK, United Kingdom; GDP, Gross Domestic Product; CPI, Consumer Prices Index

Figure 2: Information Entropy transformed (IE) UK GDP and IE UK CPI, relative to the year 2000, over the period 2000 to 2019

(a) Growth of UK GDP over the period 2000 to 2019

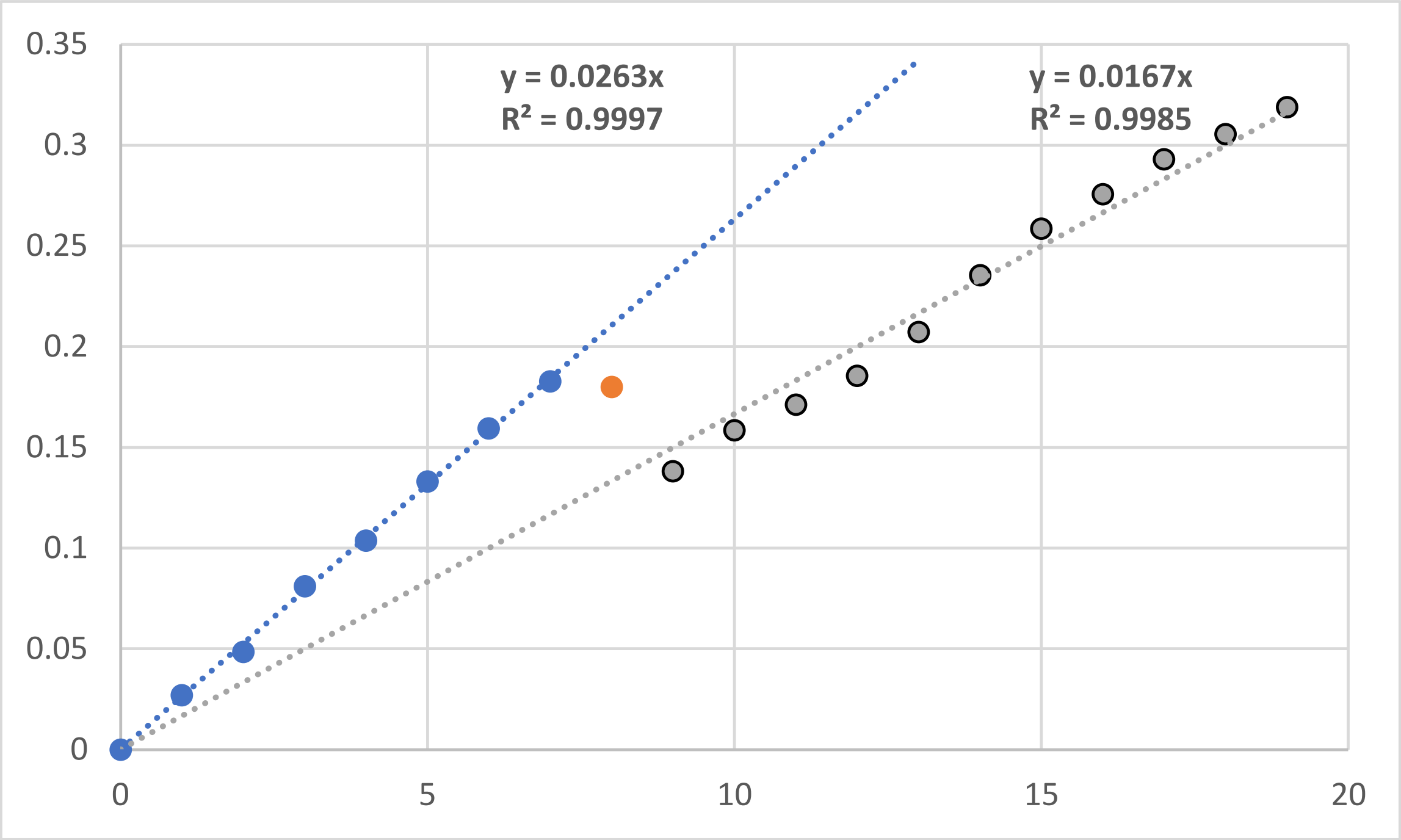

(b) Growth of UK GDP over the period 2000 to 2019

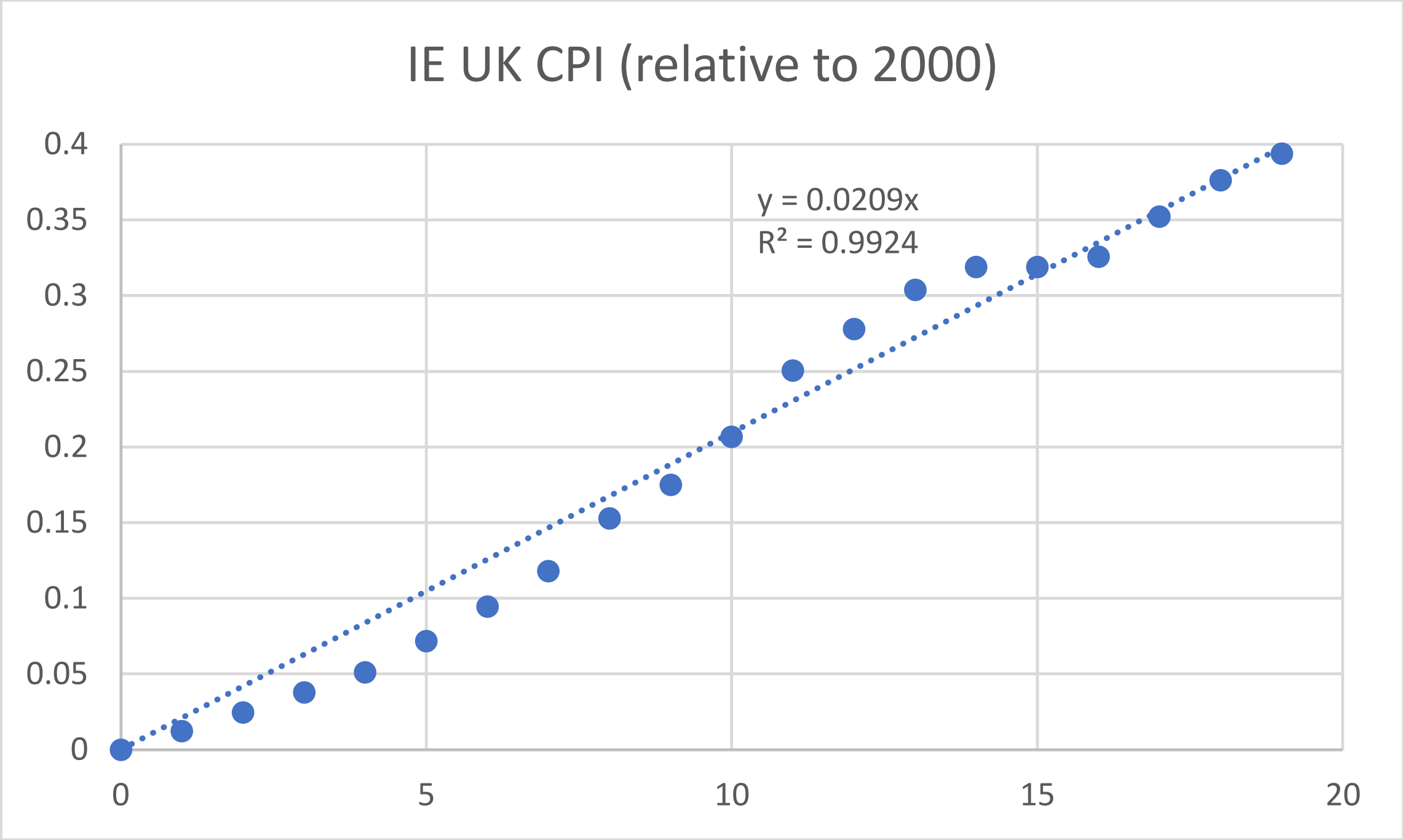

Note: IE, information entropy; UK, United Kingdom; $R^2$, coefficient of determination

## 3.2 Relationship between the growth of UK labour productivity and UK GDP per capita over the period 2000 to 2019

The IE UK GDP per capita, relative to the year 2000, over the time period (1) 2000-2007; (2) 2008-2013; (3) 2014-2019 is given in Figure 3. The rate of growth of the UK economy over the period 2000 to 2019 can be described in terms of three distinct phases:

(1) 2000 to 2007, with an annual average growth rate in GDP per capita of approximately 2.1%
(2) 2008 to 2013 – the impact of the international financial crisis, its immediate aftermath, and period of recovery
(3) 2014 to 2019, with an annual average growth rate in GDP per capita of approximately 1.0%

The annual average growth rate in UK GDP per capita over the period 2014 to 2019 was more than 50% lower than that achieved over the period 2000 to 2007.

IE UK labour productivity is compared to IE UK GDP per capita, relative to the year 2000, over the time period 2000 to 2019, in Figure 4. It can be seen that IE UK labour productivity and IE UK GDP per capita tend to increase in parallel over the periods: (1) 2000 to 2007, and (2) 2014 to 2019. The linear relationships between IE UK GDP per capita and IE UK productivity, relative to the year 2000, between (1) 2000 to 2007 and (2) 2014 to 2019 are quantified in Figure 5. The slopes are similar for both time periods being approximate 1.24 for the time period 2000 to 2007, and 1.14 for the time period 2014 to 2019 (Figure 5).

Figure 3: IE UK GDP per capita, relative to the year 2000, over the time period **(1) 2000-2007**; **(2) 2008-2013**; **(3) 2014-2019**

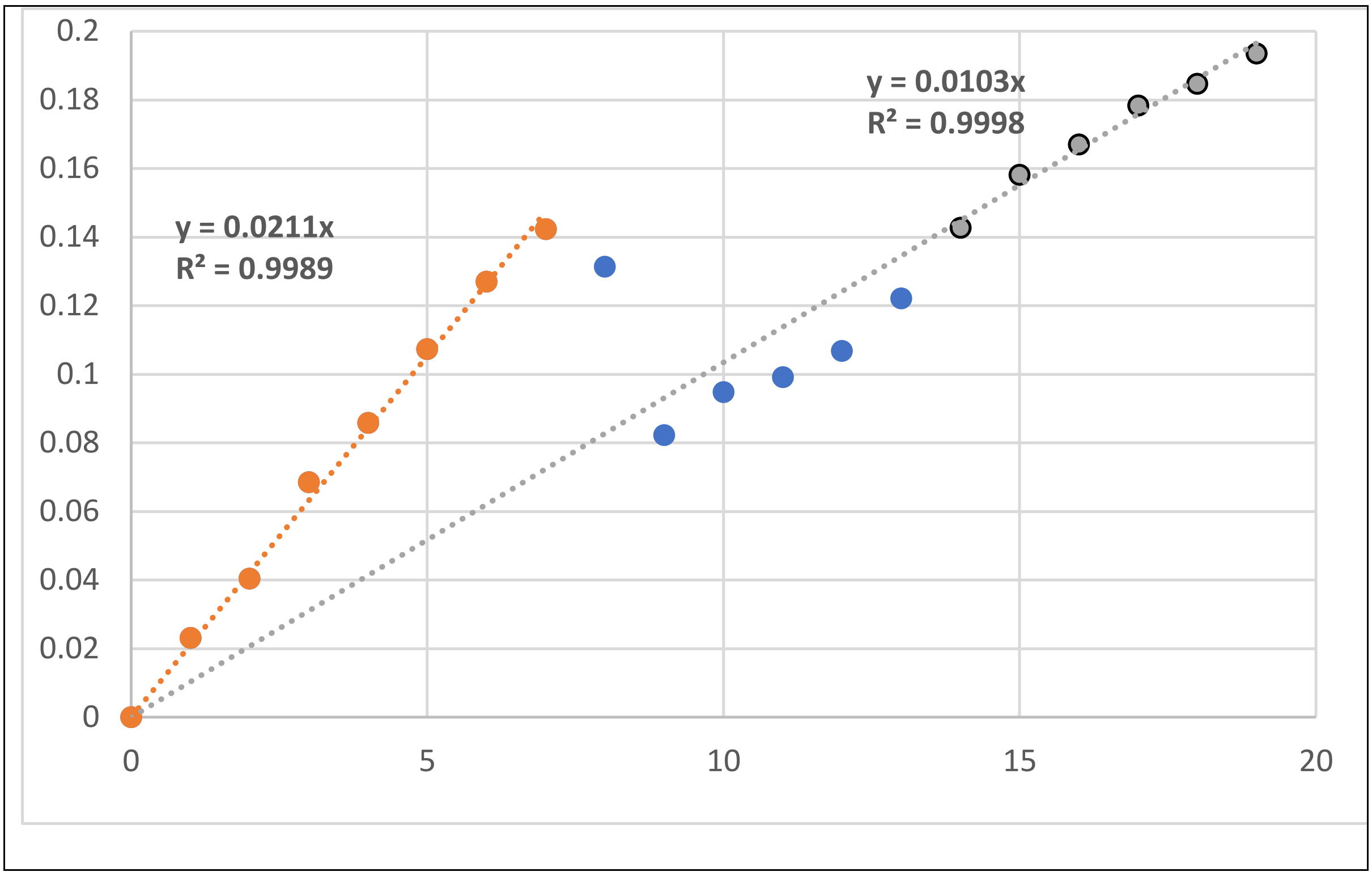

Note: IE, information entropy; UK, United Kingdom; GDP, Gross Domestic Product; $R^2$, coefficient of determination

Figure 4: IE UK labour productivity compared to IE UK GDP per capita, relative to the year 2000, over the time period 2000 to 2019

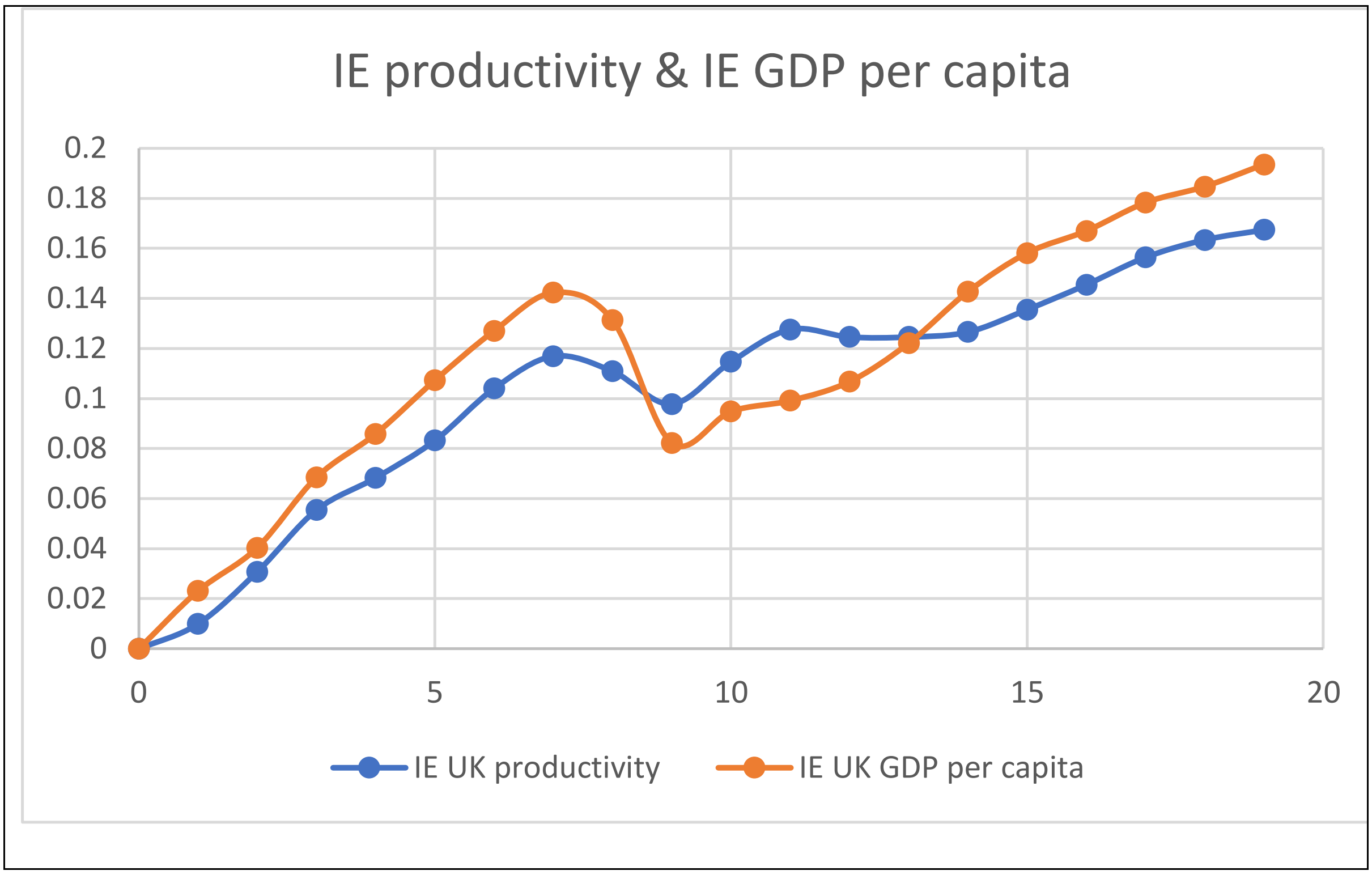

Note: IE, information entropy; UK, United Kingdom

Figure 5: Relationship between IE UK GDP per capita and IE UK productivity, relative to the year 2000, between (1) 2000 to 2007 and between (2) 2014 to 2019

(1) 2000 to 2007

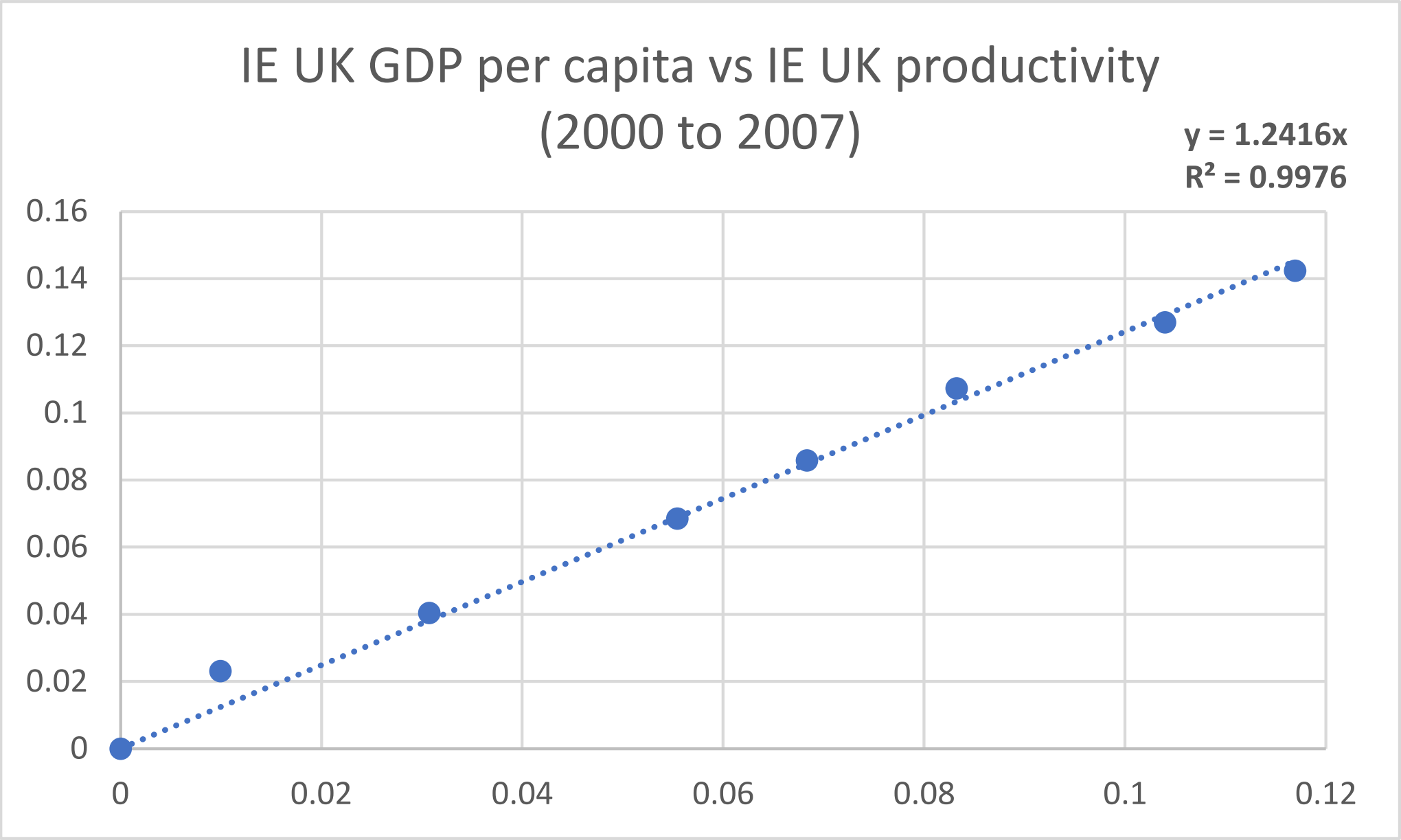

(2) 2014 to 2019

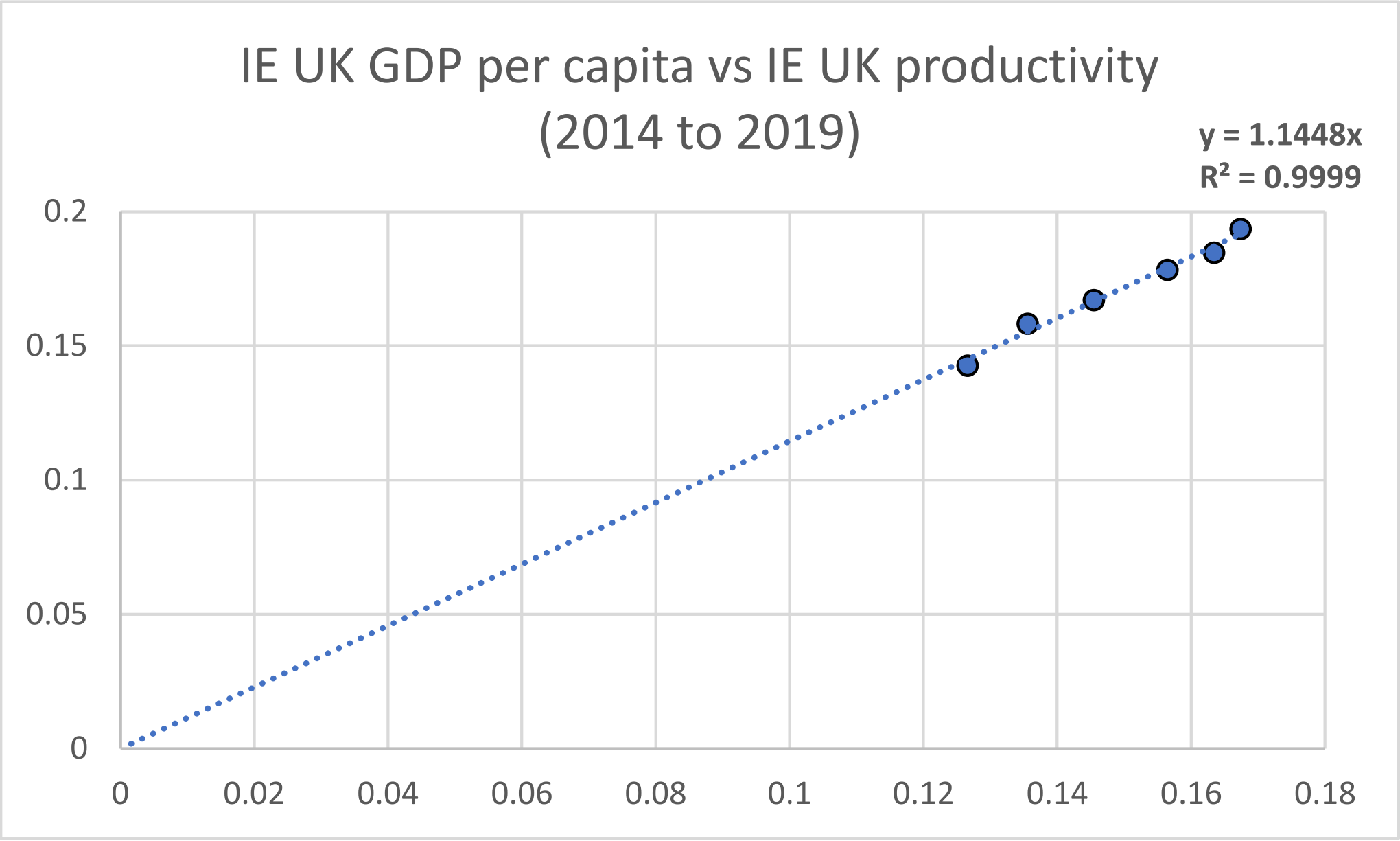

Note: IE, information entropy; UK, United Kingdom; $R^2$, coefficient of determination

## 3.3 Relationship between the growth of average UK real wages and UK labour productivity and over the period 2000 to 2019

IE average UK real wages is compared to IE UK labour productivity, relative to the year 2000, over the time period 2000 to 2019, in Figure 6. It can be seen that IE average UK real wages and IE UK labour productivity tend to increase in parallel over the periods: (1) 2000 to 2007, and (2) 2014 to 2019. The linear relationships between IE average UK real wages and IE UK productivity, relative to the year 2000, between (1) 2000 to 2007 and (2) 2014 to 2019 are quantified in Figure 7. The slope was approximately 1.51 for the time period 2000 to 2007, and 1.05 for the time period 2014 to 2019 (Figure 7), which is approximately 30% smaller than for the period 2000 to 2007.

Figure 6: IE UK average wages compared to IE UK labour productivity, relative to the year 2000, over the time period 2000 to 2019

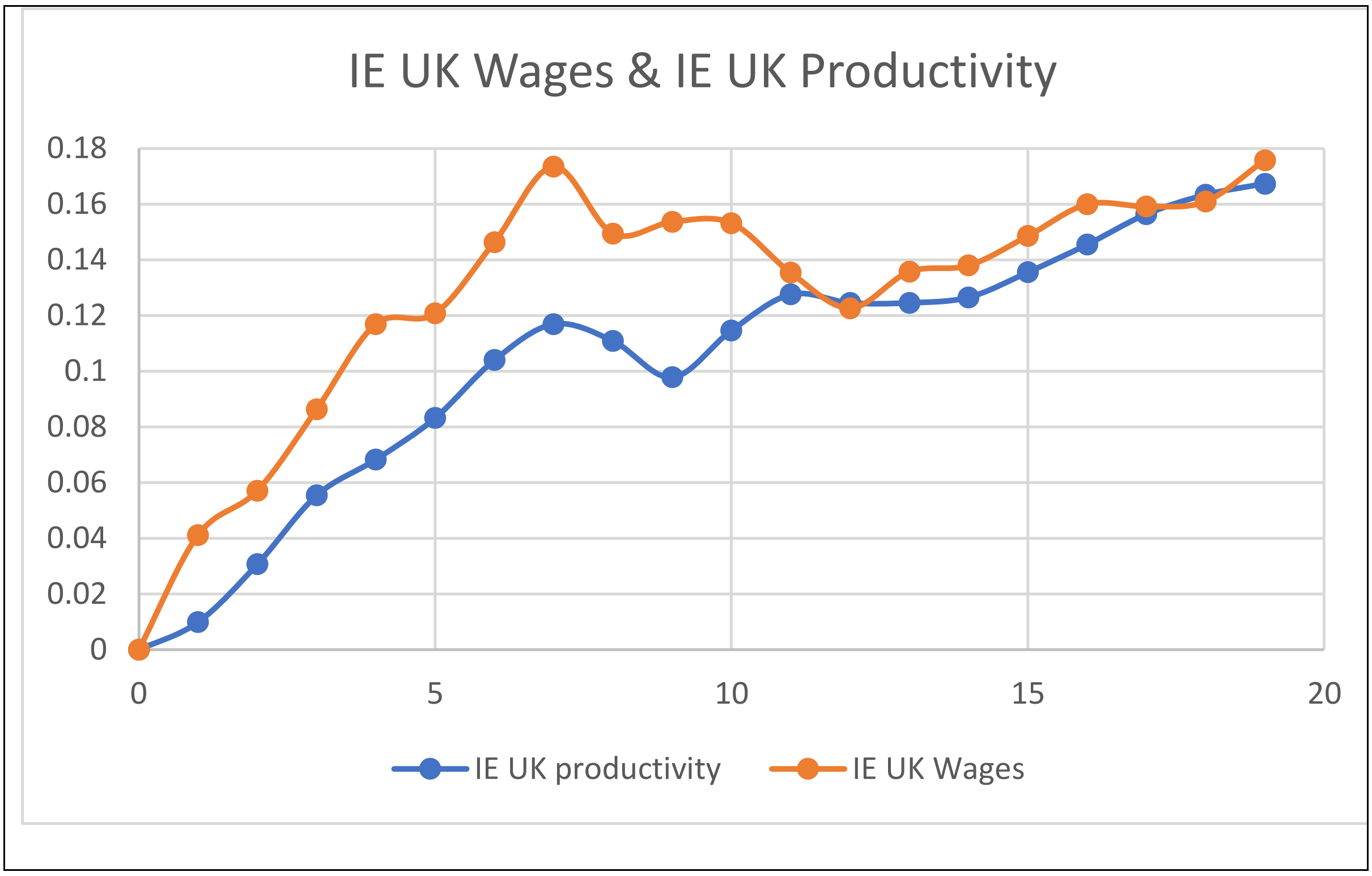

Note: IE, information entropy; UK, United Kingdom

Figure 7: Relationship between IE UK average wages and IE UK productivity between, relative to the year 2000, between (1) 2000 to 2007 and between (2) 2014 to 2019

(1) 2000 to 2007

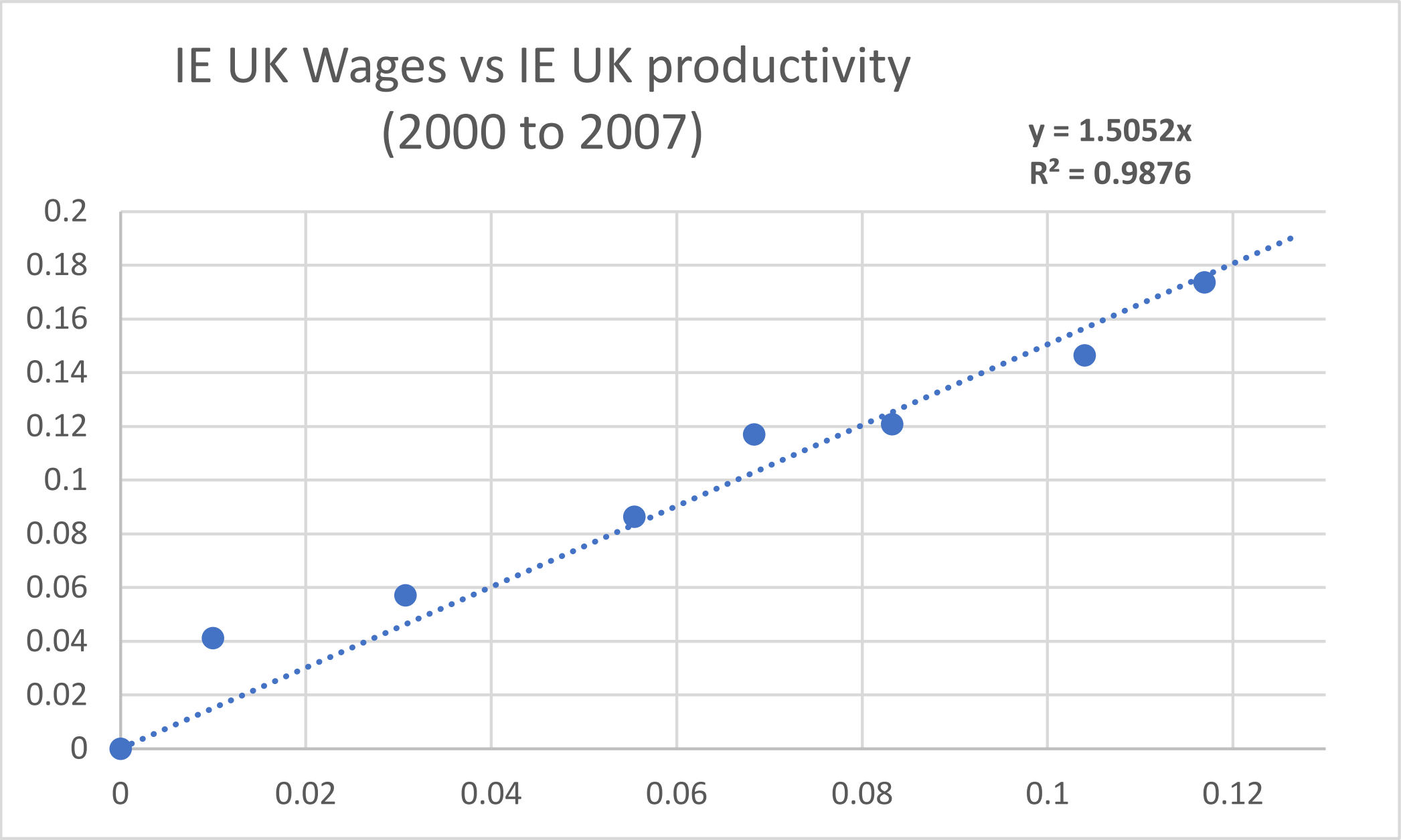

(2) 2014 to 2019

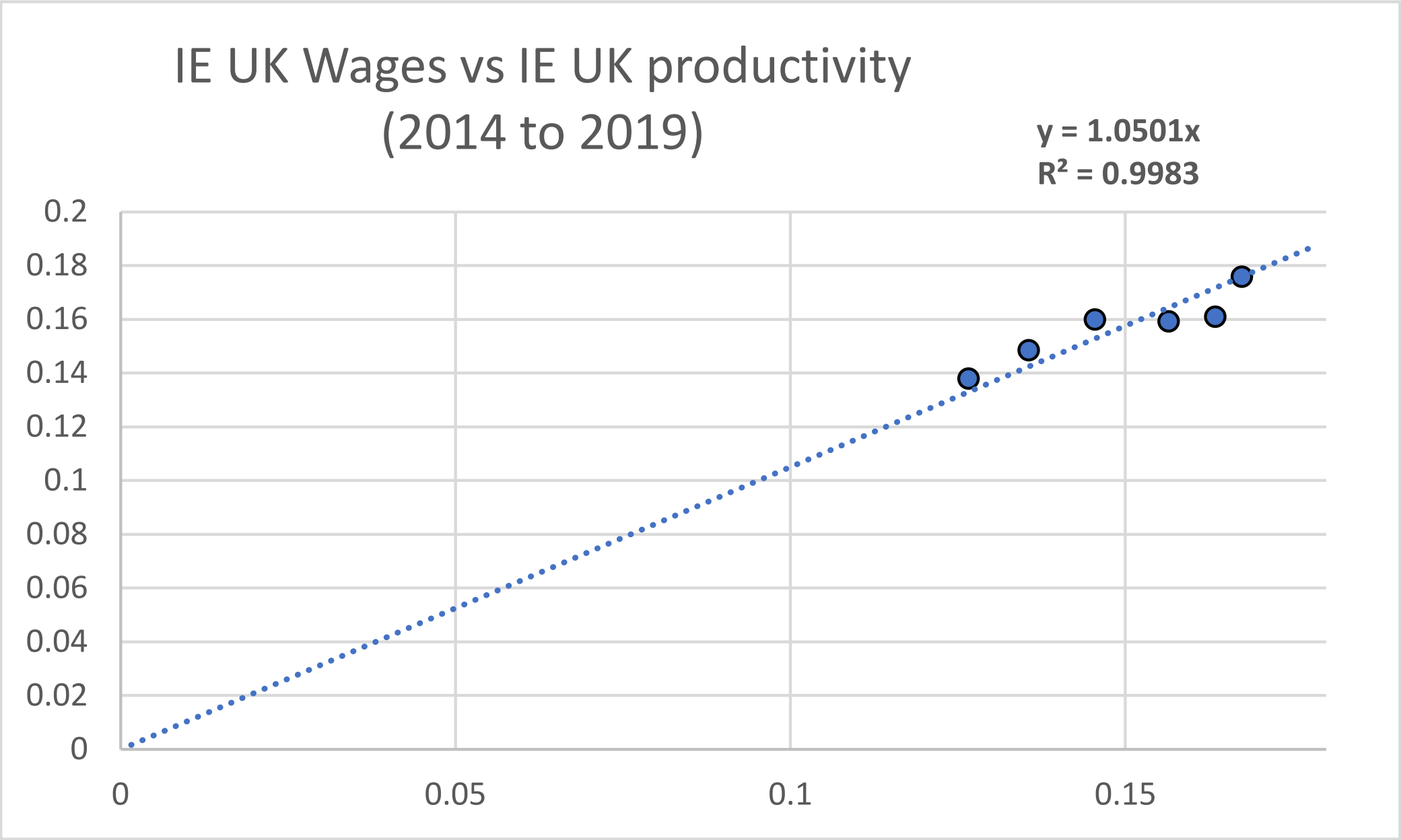

Note: IE, information entropy; UK, United Kingdom; $R^2$, coefficient of determination

# 4. Discussion

## 4.1 Study findings

While throughout the period 2000 to 2019, UK CPI was maintained, on average, at approximately 2.1% per annum, the growth of the UK economy over the period 2000 to 2019 can be described in terms of three distinct phases:

(1) 2000 to 2007 – strong sustained economic growth

(2) 2008 to 2013 – the impact of the international financial crisis, its immediate aftermath, and period of recovery

(3) 2014 to 2019 – weak sustained economic growth

The key determinant of the UK economic performance over this period would appear to the annual rate of growth in labour productivity. It was closely related to the annual rate of growth in GDP per capita, and it was significantly weaker in the period 2014 to 2019 compared to the period 2000 to 2007. This also corresponded with a weaker rate of growth in average real wages over the period 2014 to 2019 compared to the period 2000 to 2007.

These findings are supported by macroeconomic literature and theory. Labour productivity is one of the most important factors contributing to economic growth [11]. Economic theory holds that at the aggregate level, the growth of real wages are determined by labour productivity growth, and only labour productivity growth can raise living standards in the long term [12], but this does not mean that wage growth will be equally distributed across different categories of workers [13].

## 4.2 Implications for policy makers

Theoretically, the main determinants of labour productivity are physical capital, human capital, and technological change [14]. Therefore, the ways that governments and companies can improve labour productivity is by means of investment in [15]:

(1) Physical capital: capital goods, including infrastructure

(2) Education and training: opportunities for workers to upgrade their skills

(3) New technologies: hardware (e.g., robotics) and software; innovation [14]

The link between labour productivity and investment in the UK over the period 2000 to 2019 was investigated. In keeping with the rest of this paper this was done over the three separate time periods: (1) 2000 to 2007; (2) 2008 to 2014; (3) 2015 to 2019, with a focus on the periods pre-international financial crisis (2000 to 2007) and following recovery from it (2015 to 2019). The source of UK investment was Total Gross Fixed Capital Formation (GFCF) [16] over the period 2000 to 2019, obtained from the UK ONS (NPQT, time series data) [17].

IE UK labour productivity is compared to IE UK investment, relative to the year 2000, over the time period 2000 to 2019, in Figure 8. It can be seen that UK labour productivity and IE UK investment tend to both increase over the periods: (1) 2000 to 2007, and (2) 2014 to 2019. The linear relationships between IE UK labour productivity and IE UK investment, relative to the year 2000, between (1) 2000 to 2007 and (2) 2014 to 2019 are quantified in Figure 9. The slope was approximately 0.65 for the time period 2000 to 2007, and 0.54 for the time period 2014 to 2019 (Figure 9), which is approximately 17% smaller than for the period 2000 to 2007.

Figure 8: IE UK labour productivity compared to IE UK investment, relative to the year 2000, over the time period 2000 to 2019

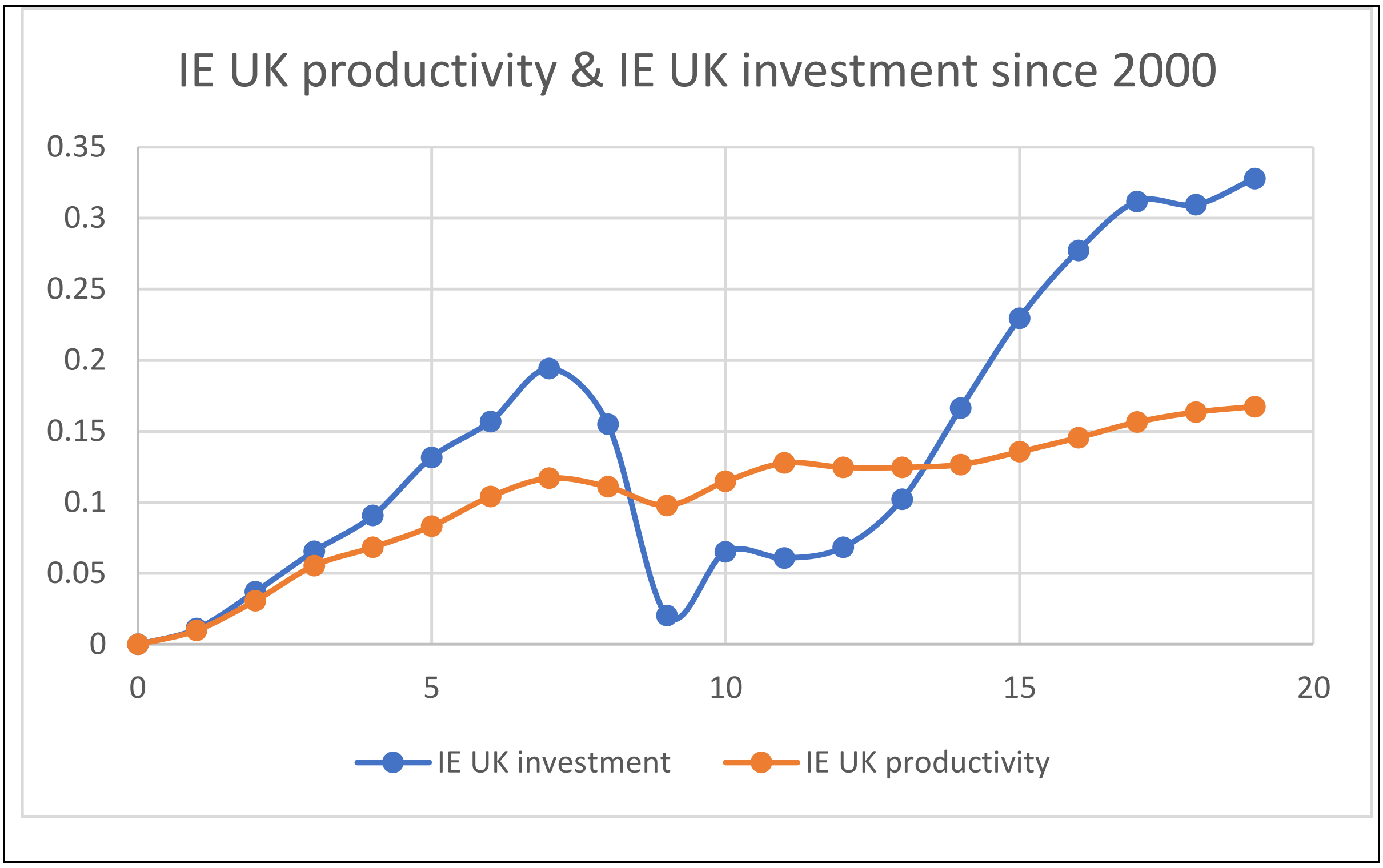

Note: IE, information entropy; UK, United Kingdom

Figure 9: Relationship between IE labour productivity and IE UK investment between, relative to the year 2000, between (1) 2000 to 2007 and between (2) 2014 to 2019

(1) 2000 to 2007

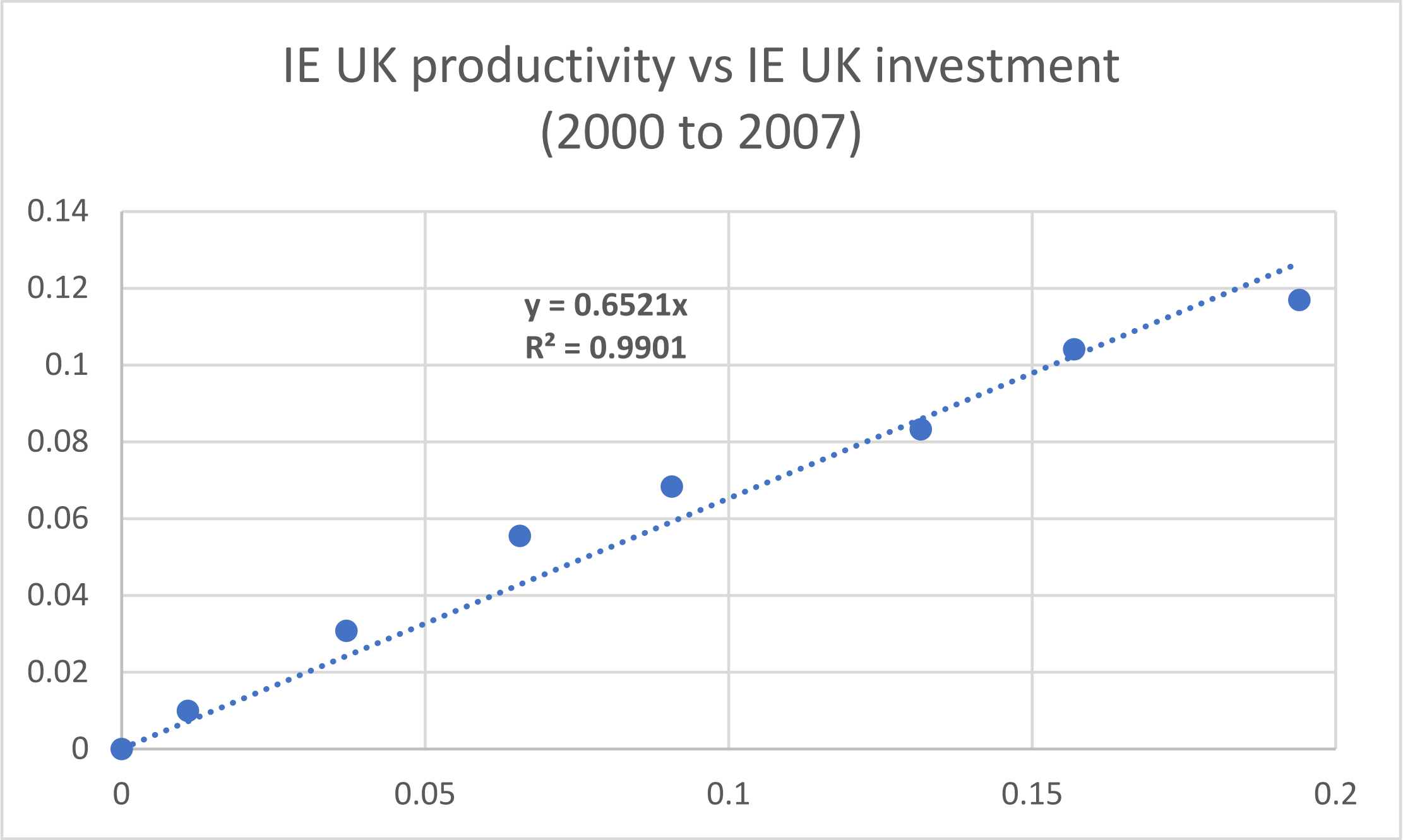

(2) 2014 to 2019

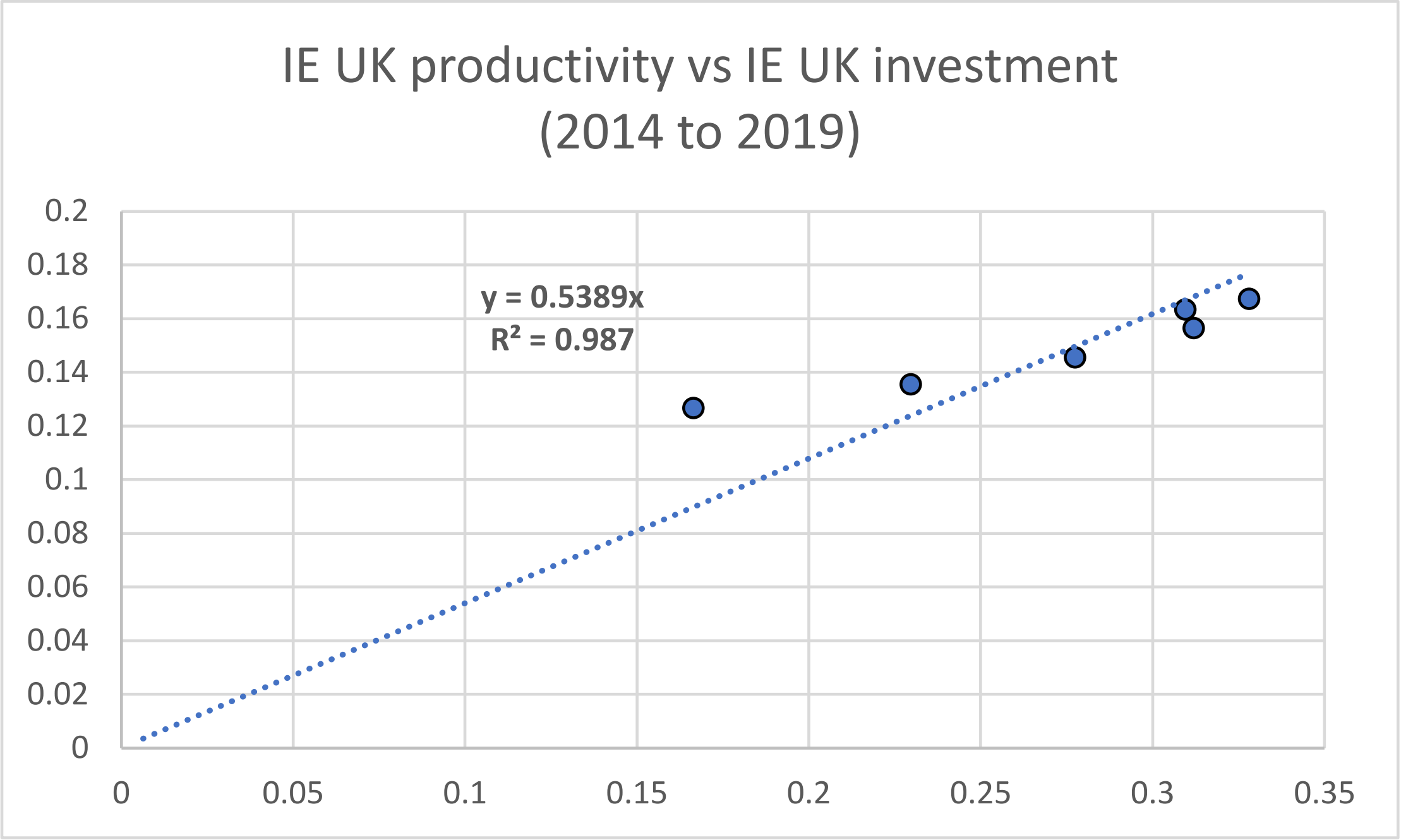

Note: IE, information entropy; UK, United Kingdom; $R^2$, coefficient of determination

The relationship between growth in UK labour productivity and the growth in UK investment, as measured by Total GFCF, may not have "returned normal" following recovery from the international financial crisis until after 2014 (Figure 9). Furthermore, there may be a more appropriate measure of UK investment for the purposes of investigating labour productivity than Total GFCF, but that was outside the scope of this paper.

Using the quantitative linear relationships (on the log scale) outlined in this paper between:

(1) Growth of UK labour productivity with growth of UK investment

(2) Growth of UK GDP per capita with growth of UK labour productivity, and

(3) Growth of the UK population over the period 2000 to 2019 [18]

it is possible to predict the growth of UK GDP over this time period and to compare the predictions with the actual observed GDP time series. These are given in Figure 10. It can be seen that the predictive accuracy of the methodology, over the periods 2000 to 2007 and 2014 to 2019, is approximate 99.8% (1/1.0023)

While the results given in this paper are specific to the UK over the time period 2000 to 2019, the expectation is that the methodology and approach adopted can be applied to quantifying the dynamics of any developed economy over any time period.

Figure 10: Predicted and observed UK GDP over the period 2000 to 2010, with the predictive accuracy of the methodology obtained over the periods 2000 to 2007 and 2014 to 2019

(1) Observed and predicted UK GDP (2000 to 2019)

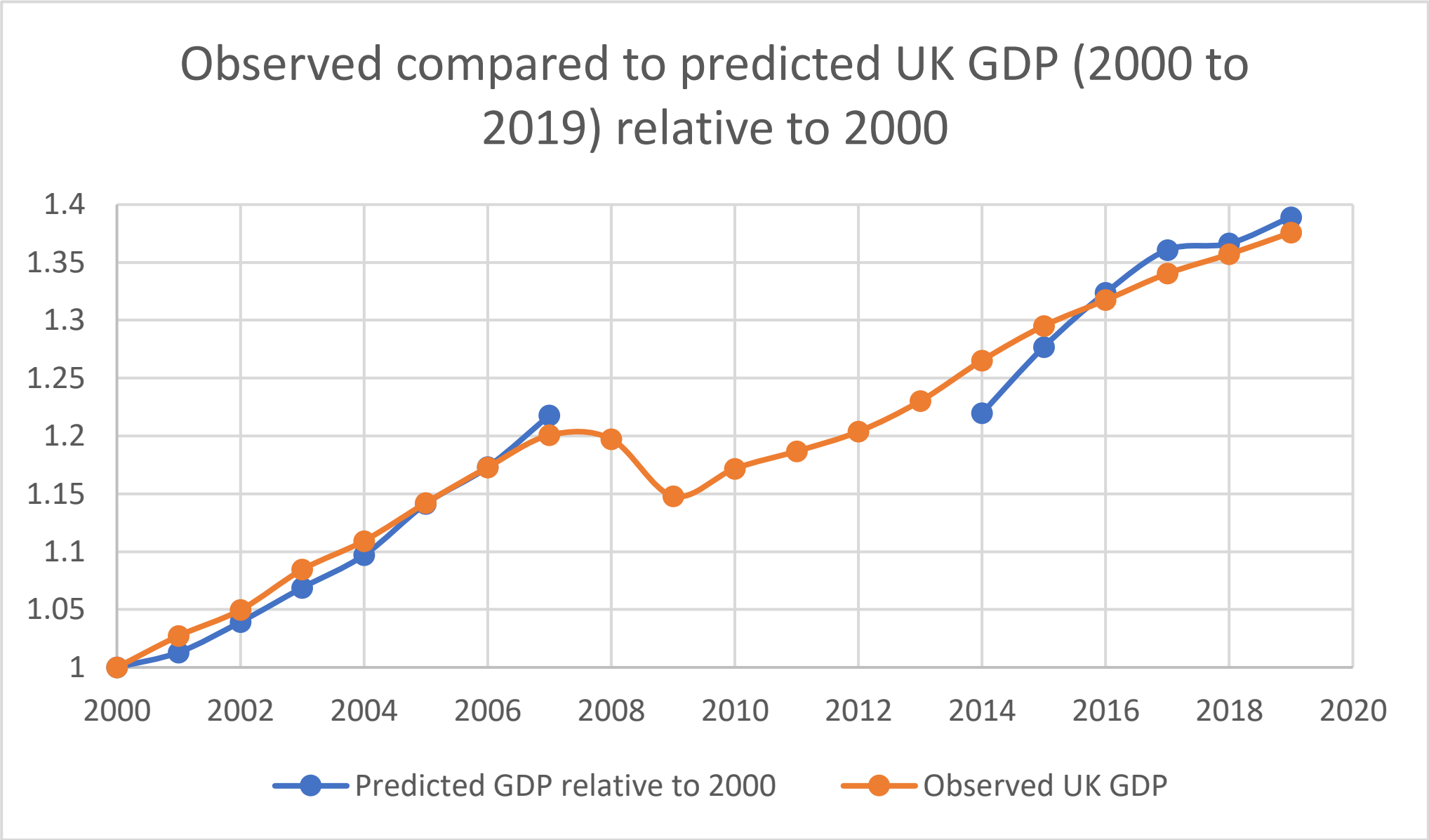

(2) Predictive accuracy over the periods 2000 to 2007 and 2014 to 2019

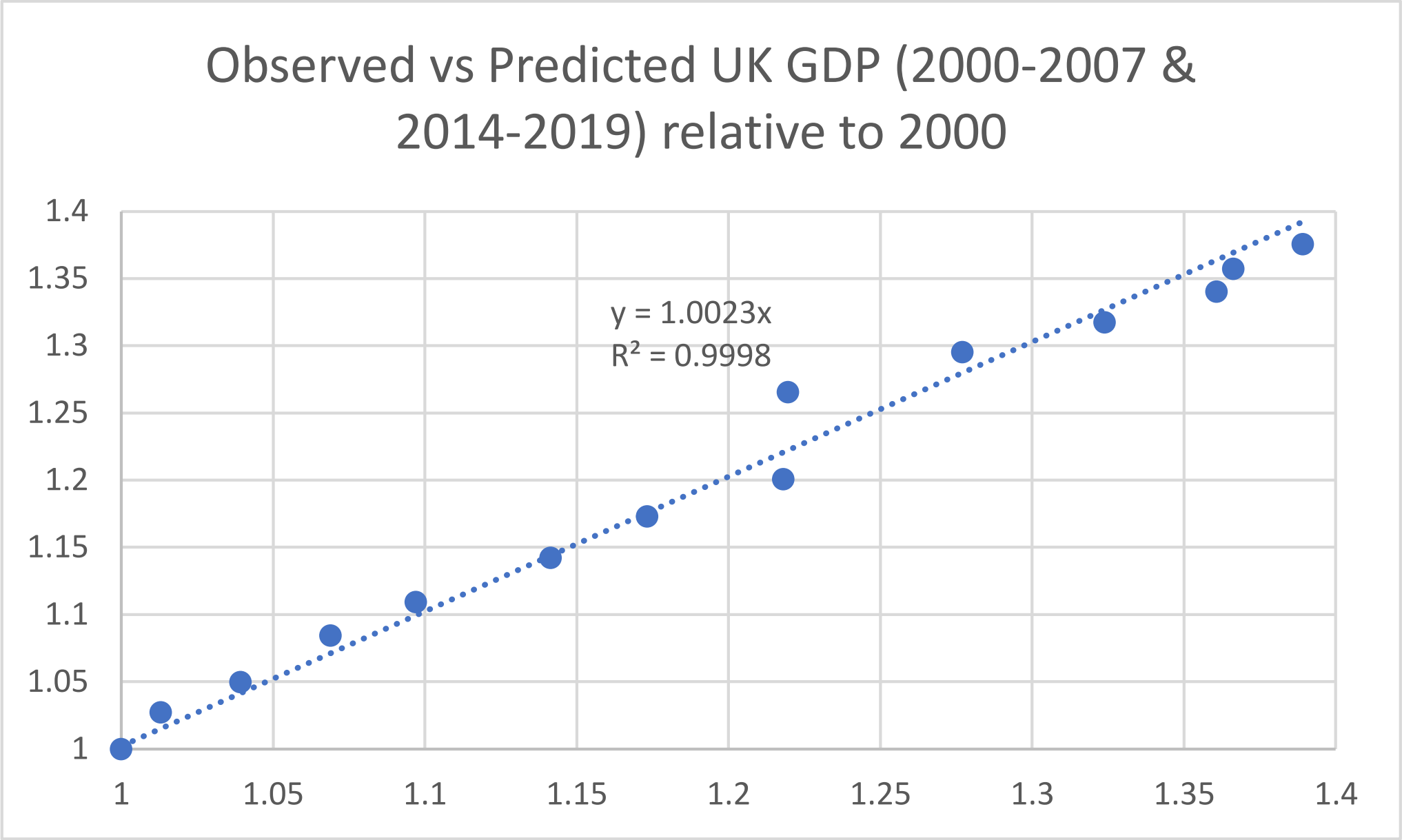

Note: IE, information entropy; UK, United Kingdom; $R^2$, coefficient of determination

## 5. Conclusion

This economic evaluation would suggest that more rapid UK economic growth can be achieved by sustained investment in measures that enhance labour productivity, with the expectation that a sustained improvement in labour productivity will increase the annual rate of growth of UK GDP per capita and average real wages. While the results given in this paper are specific to the UK over the time period 2000 to 2019, the expectation is that the methodology and approach adopted can be applied to quantifying the dynamics of any developed economy over any time period.

## Supplementary materials

There are none.

## Acknowledgements

No financial support was received for any aspect of this research.